\documentclass[twocolumn,english]{IEEEtran}
\usepackage[T1]{fontenc}
\usepackage{amsthm}
\usepackage{amssymb}
\usepackage{graphicx}

\makeatletter
\theoremstyle{plain}
\newtheorem{thm}{\protect\theoremname}
\theoremstyle{plain}
\newtheorem{cor}[thm]{\protect\corollaryname}
\theoremstyle{definition}
\newtheorem{defn}[thm]{\protect\definitionname}
\theoremstyle{plain}
\newtheorem{lem}[thm]{\protect\lemmaname}

\usepackage[a4paper, bindingoffset=0.0in,left=0.625in,right=0.625in,top=0.75in,bottom=1in, footskip=.25in]{geometry}
\usepackage{blindtext}
\usepackage{footnote}
\makesavenoteenv{tabular}

\usepackage{lettrine}
\usepackage{hyperref}

\usepackage{amsthm}

\usepackage{amsmath}

\usepackage{algorithmic}
\usepackage{float}
\newfloat{algorithm}{t}{lop}

\usepackage{cite}
\usepackage{url}
\usepackage[small]{caption}

\newlength{\figwidth}
\ifCLASSOPTIONonecolumn
\else
\fi

\usepackage{graphicx}
\usepackage{epstopdf}
\usepackage{amssymb}
\usepackage{subfig}
\usepackage{amsmath}
\usepackage[misc,geometry]{ifsym}

\makeatother

\usepackage{babel}
\providecommand{\corollaryname}{Corollary}
\providecommand{\definitionname}{Definition}
\providecommand{\lemmaname}{Lemma}
\providecommand{\theoremname}{Theorem}

\begin{document}

\title{Throughput Analysis for Wireless Networks\\ with Full-Duplex Radios}

\author{Zhen Tong and Martin Haenggi\\
Department of Electrical Engineering \\
University of Notre Dame, 
              Notre Dame, IN 46556, USA\\
              E-mail: {\{ztong1,mhaenggi\}@nd.edu}}
\maketitle

\begin{abstract}
This paper investigates the throughput for wireless network with full-duplex radios using stochastic geometry. Full-duplex (FD) radios can
exchange data simultaneously with each other. On the other hand, the
downside of FD transmission is that it will inevitably cause extra
interference to the network compared to half-duplex (HD) transmission.
In this paper, we focus on a wireless network of nodes with both HD
and FD capabilities and derive and optimize the throughput
in such a network. Our analytical result shows that if the network is adapting an ALOHA
protocol, the maximal throughput is always achieved by scheduling all concurrently transmitting nodes
to work in FD mode instead of a mixed FD/HD mode or HD mode regardless
of the network configurations. Moreover, the throughput gain of using FD transmission over HD transmission is analytically lower
and upper bounded.
\end{abstract}

\section{Introduction}

Traditionally, radio transceivers are subject to a  HD constraint
because of the crosstalk between the transmit and receive chains.
The self-interference caused by the transmitter at the receiver if using
FD transmission overwhelms the desired received signal from the partner
node since it is  much stronger than the desired received
signal. Therefore, current radios all use orthogonal signaling dimensions, i.e., time division
duplexing (TDD) or frequency division duplexing (FDD), to achieve bidirectional
communication. 

FD communication can potentially double the throughput if the self-interference
can be well managed. FD radios have been successfully implemented in the industrial, scientific and medical (ISM) radio bands in a laboratory environment in the past few years~\cite{ChoJai10Mobicom, DuaSab10Asilomar, JaiCho11Mobicom, SahPat11X}.  Key
to the success are novel analog and digital self-interference cancellation
techniques as well as spatially separated transmit and receive antennas.
A FD system with only one antenna has also been implemented in \cite{Knox12WAMICON}
by using specially designed circulator. In general, the main idea
is to let the receive chain of a node remove the self-interference
caused by the known signal from its transmit chain, so that reception
can be concurrent with transmission. A novel signaling technique
was proposed in \cite{Guo10Allerton} to achieve virtual FD with applications
in neighbor discovery \cite{Guo2-13} and mutual broadcasting \cite{Guo13}

From a theoretical perspective, the two-way transmission capacity
of wireless ad hoc networks has been studied in \cite{Vaze11} for
a FDD model. A FD cellular system has been analyzed in \cite{Goyal2013}
where the throughput gain has been illustrated via extensive simulation
for a cellular system with FD
base station and HD mobile users. The throughput gain of
single cell MIMO wireless systems with FD radios has been
quantified in \cite{Barghi12}. A capacity analysis of FD and HD transmissions with bounded
radio resources has been presented in \cite{Aggarwal12ITW} with focus
on a single-link system. \cite{Ju12,Kim14} evaluate the capacity of FD ad hoc networks and alleviate the capacity
degradation due to the extra interference of FD by using beamforming and ARQ protocol respectively. Both
capacity analyses in \cite{Ju12,Kim14} are based on the approximation
that the distances between the desired receiver and the interfering
pair are the same.

In this paper, the impacts of FD transmission on the network throughput
are explored. On the one hand, FD transmission allows bidirectional communication
between two nodes simultaneously and therefore potentially doubles the throughput.
On the other hand, the extra interference caused by FD transmissions can
degrade the throughput gain over HD, which makes it unclear that FD
can actually outperform HD for a given network configuration.
This paper utilizes the powerful analytical tools from stochastic geometry
to study the throughput performance of a wireless network of nodes with both FD and HD capabilities. Our results analytically
show that for an ALOHA MAC protocol, FD always outperforms HD in terms of throughput if perfect self-interference cancellation is assumed.
This result holds for arbitrary node densities, path loss
exponents, link distances and SINR regimes. 


\section{Network Model\label{sec:Network-Model}}

Consider an independently marked Poisson point process (PPP) \cite{Haenggi12book}
$\hat{\Phi}=\left\{ \left(x_{i},m(x_{i}),s_{x_{i}}\right)\right\} $
on $\mathbb{R}^{2}\times\mathbb{R}^{2}\times\left\{ 0,1,2\right\} $
where $\Phi=\left\{ x_{i}\right\} $ is a PPP with density $\lambda$
and $m(x_{i})$ and $s_{x_{i}}$ are the marks of point
$x_{i}$. The mark $m(x_{i})$ defines the node that $x_{i}$
 communicates with. Here, we fix $\left\Vert x-m(x)\right\Vert =R$, $\forall x\in\Phi$,
i.e., $R$ is the distance of all links. Therefore, $m(x_{i})$ can also be written as $m(x_{i})=x_{i}+R\left(\cos\varphi_{i},\sin\varphi_{i}\right)$,
where the $\varphi_{i}$s are independent and uniformly distributed on $\left[0,2\pi\right]$. The link distance $R$ can also be random and the main conclusion in this paper is not affected since we can always derive the results by first conditioning on $R$ and then averaging over $R$. We define $m(\Phi)=\{m(x): x\in\Phi\}$, which is also a PPP of density $\lambda$.
The mark $s_{x_{i}}$ indicates the independently chosen state of the link that consists
of $x_{i}$ and $m(x_{i})$: $s_{x_{i}}=0$ means the link is silent,
$s_{x_{i}}=1$ means the link is in HD mode, and $s_{x_{i}}=2$ means in FD mode. HD means that in a given
time slot the transmission is unidirectional, i.e., only from $x_{i}$
to $m(x_{i})$, while FD means that $x_{i}$ and $m(x_{i})$
are transmitting to each other. Therefore, for any link there are three
states: silence, HD and FD. Assume that a link is in the state of
silence with probability $p_{0}$, HD with probability $p_{1}$ and
FD with probability $p_{2}$, where $p_{0}+p_{1}+p_{2}=1$. $p_{1}$
and $p_{2}$ are the medium access probabilities (MAPs) for HD and
FD modes respectively. As a result, $\Phi=\bigcup_{i=0}^{2}\Phi_{\left[i\right]}$,
where $\Phi_{\left[i\right]}=\left\{ x\in\Phi: s_{x}=i\right\} $ with density $\lambda p_{i}$
and $i\in\left\{ 0,1,2\right\}$. From the coloring theorem \cite[page 53]{kingman-poisson-processes},
these three node sets $\Phi_{\left[i\right]}$ are independent from
each other. 

\begin{figure}[h]
\vspace{-3mm}
\begin{centering}
\includegraphics[width=\figwidth]{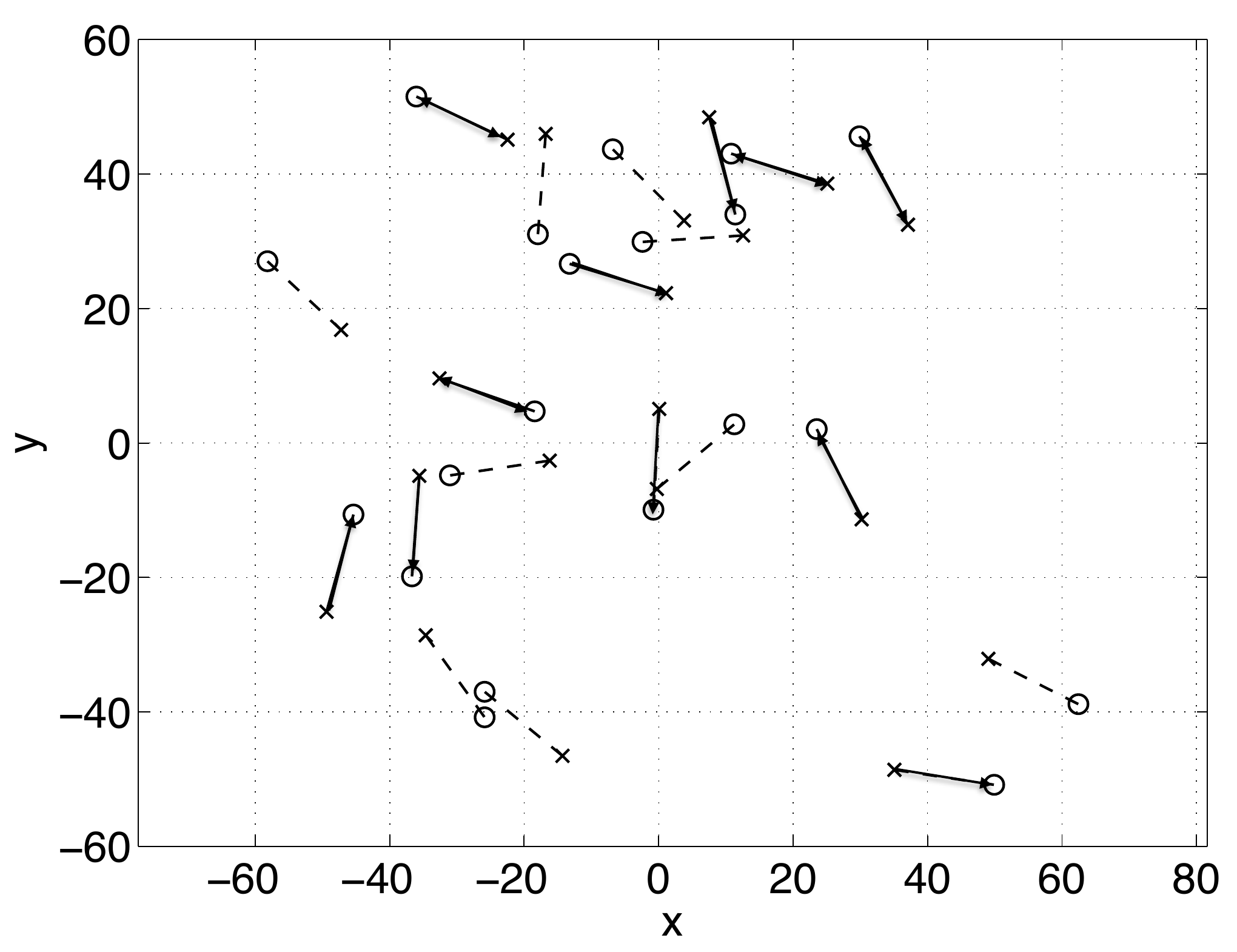}
\par\end{centering}
\caption{An example of the class of wireless networks considered in this paper
where the dashed lines indicate the link is being silent, the arrows
mean the link is in HD mode and the double arrows in FD mode. The x's form $\Phi$ while the o's form $m(\Phi)$.}
\label{Fig:0}
\end{figure}

The marked point process $\hat{\Phi}$ can be used to
model a wireless network of nodes with both FD and HD capabilities.
The self-interference in the FD links is assumed to be cancelled
perfectly. In the following, we will use this model to study the performance
of wireless networks with FD radios. An example of such a wireless
network is illustrated in Figure \ref{Fig:0}. 
In this network setup, consider the SIR model where a transmission
attempt from $x$ to $y$ is considered successful if 
\[
\mbox{SIR}_{y}=\frac{h_{xy}l(x,y)}{\sum_{z\in\tilde{\Phi}\backslash\left\{ x\right\} }h_{yz}l(z,y)}>\theta,
\]
where $\tilde{\Phi}$ is the set of transmitting nodes in a given time slot, $\theta$ is the SIR threshold,
and $h_{xy}$ and $h_{zy}$ are the fading power coefficients with
mean $1$ from the desired transmitter $x$ and the interferer $z$
to $y$ respectively. The transmit powers are fixed to $1$. We focus on the Rayleigh fading case for both
the desired link and interferers. The path loss function $l(x,y)$
between node $x$ and $y$ considered is $l\left(x,y\right)=\left\Vert x-y\right\Vert ^{-\alpha}$
where $\alpha>2$ is the path-loss exponent. If $y$ is at the origin, the index $y$ will be omitted. Also, we
define a given set of system parameters $\left(\lambda,\theta,R,\alpha\right)$
as one \textit{network configuration}. We will show that
some conclusions hold regardless of the network configuration.

\section{Success Probability\label{sec:Success-Probability}}

Our first metric of interest is the success probability, defined as
\vspace{0mm}
\begin{equation}
p_{s}=\mathbb{P}\left(\mbox{SIR}_{y}>\theta\right),\label{eq:ps}
\end{equation}
which is also the ccdf of the SIR. Without changing the distribution of the point process, assume that
the receiver $y$ is at the origin. This implies there is a transmitter at fixed distance
$R$ from the origin. The success probability plays an
important role in determining the throughput as will be described
in the following section.

The following theorem gives the success probability of the FD/HD-mixed
wireless network modeled by the marked PPP:
\begin{thm}
In a wireless network described by the marked PPP $\hat{\Phi}$, the success
probability defined in (\ref{eq:ps}) is given by 
\begin{equation}
p_{s}=\exp\left(-\lambda p_{1}G(\theta R^{\alpha},\alpha)\right)\exp\left(-\lambda p_{2}F(\theta R^{\alpha},\alpha,R)\right),\label{eq:ps-2}
\end{equation}
where $G(s,\alpha)=\frac{\pi^{2}\delta s^{\delta}}{\sin\left(\pi\delta\right)}$
with $\delta\triangleq2/\alpha$ and
\ifCLASSOPTIONonecolumn
\[
F(s,\alpha,R)=\int_{0}^{\infty}\left(1-\frac{1}{1+sr^{-\alpha}}\int_{0}^{2\pi}\frac{d\varphi}{1+s\left(r^{2}+R^{2}+2rR\cos\varphi\right)^{-\alpha/2}}\right)rdr.
\]
\else
\[F(s,\alpha,R)= \int_{0}^{\infty}\left(1-\frac{1}{1+sr^{-\alpha}}K(s,r,R,\alpha)\right)rdr\]
with $K(s,r,R,\alpha)=\int_{0}^{2\pi}\frac{d\varphi}{1+s\left(r^{2}+R^{2}+2rR\cos\varphi\right)^{-\alpha/2}}.$
\fi
\end{thm}
\begin{IEEEproof}
With Rayleigh fading, the desired signal strength $S$ at the
receiver at the origin is exponential, i.e., $S=hR^{-\alpha}$. The interference $I$ consists of two parts: the
interference from the HD nodes $\Phi_{\left[1\right]}$ and the interference from the FD nodes $\Phi_{\left[2\right]}$. It can be expressed in the following form:
\[
I=\sum_{x\in\Phi_{\left[1\right]}}h_{x}l(x)+\sum_{x\in\Phi_{\left[2\right]}}\left(h_{x}l(x)+h_{m(x)}l(m(x))\right).
\]
The Laplace transform of the interference follows as
\ifCLASSOPTIONonecolumn
\begin{eqnarray}
L_{I}\left(s\right) & = & \mathbb{E}e^{-s\left(\sum_{x\in\Phi_{\left[1\right]}}h_{x}l(x)+\sum_{x\in\Phi_{\left[2\right]}}\left(h_{x}l(x)+h_{m(x)}l(m(x))\right)\right)}\nonumber \\
 & = & \mathbb{E}\left(\prod_{x\in\Phi_{\left[1\right]}}e^{-sh_{x}l(x)}\prod_{x\in\Phi_{\left[2\right]}}e^{-s\left(h_{x}l(x)+h_{m(x)}l(m(x))\right)}\right)\nonumber \\
 & \overset{\left(a\right)}{=} & \mathbb{E}\left(\prod_{x\in\Phi_{\left[1\right]}}e^{-sh_{x}l(x)}\right)\mathbb{E}\left(\prod_{x\in\Phi_{\left[2\right]}}e^{-s\left(h_{x}l(x)+h_{m(x)}l(m(x))\right)}\right),\label{eq:Two}
\end{eqnarray}
where (a) comes from the fact that $\Phi_{\left[1\right]}$ and
$\Phi_{\left[2\right]}$ are independent PPPs from the coloring theorem
 \cite[page 53]{kingman-poisson-processes}. The first term in the
product of (\ref{eq:Two}) is the Laplace transform of the interference
of the PPP $\Phi_{\left[1\right]}$, given by \cite[page 103]{Haenggi12book}:
\else
\begin{align}
L_{I}\left(s\right)  
 & =  \mathbb{E}\left(\prod_{x\in\Phi_{\left[1\right]}}e^{-sh_{x}l(x)}\prod_{x\in\Phi_{\left[2\right]}}e^{-s\left(h_{x}l(x)+h_{m(x)}l(m(x))\right)}\right)\nonumber \\
 &  \overset{\left(a\right)}{=}  \mathbb{E}\left(\prod_{x\in\Phi_{\left[1\right]}}e^{-sh_{x}l(x)}\right)\cdot\nonumber\\
 & \qquad \mathbb{E}\left(\prod_{x\in\Phi_{\left[2\right]}}e^{-s\left(h_{x}l(x)+h_{m(x)}l(m(x))\right)}\right),\label{eq:Two}
\end{align}
where (a) comes from the fact that $\Phi_{\left[1\right]}$ and
$\Phi_{\left[2\right]}$ are independent PPPs from the coloring theorem
 \cite[page 53]{kingman-poisson-processes}. The first term in the
product of (\ref{eq:Two}) is the Laplace transform of the interference
of the PPP $\Phi_{\left[1\right]}$, given by \cite[page 103]{Haenggi12book}:
\fi
\begin{align*}
L_{I_{1}}\left(s\right) & =  \mathbb{E}\left(\prod_{x\in\Phi_{\left[1\right]}}e^{-sh_{x}l(x)}\right)\\
 & =  \exp\left(-\pi\lambda p_{1}\Gamma(1+\delta)\Gamma(1-\delta\right)s^{\delta})\\
 & =  \exp\left(-\lambda p_{1}G(s,\alpha)\right),
\end{align*} where $\Gamma(\cdot)$ is the gamma function.

The second term in the product of (\ref{eq:Two}) can be written as
follows:\ifCLASSOPTIONonecolumn
\begin{eqnarray}
L_{I_{2}}\left(s\right) & = & \mathbb{E}\left(\prod_{x\in\Phi_{\left[2\right]}}e^{-s\left(h_{x}l(x)+h_{m(x)}l(m(x))\right)}\right)\nonumber \\
 & = & \mathbb{E}\left(\prod_{x\in\Phi_{\left[2\right]}}\frac{1}{1+sl(x)}\frac{1}{1+sl(m(x))}\right)\label{eq:I2}\\
 & \overset{\left(a\right)}{=} & \exp\left(-\lambda p_{2}\int_{\mathbb{R}^{2}}\left(1-\frac{1}{1+sl(x)}\frac{1}{1+sl(m(x))}\right)\right)\label{eq:I2-1}\\
 & = & \exp\left(-\lambda p_{2}\int_{0}^{\infty}\left(1-\frac{1}{1+sr^{-\alpha}}\int_{0}^{2\pi}\frac{d\varphi}{1+s\left(r^{2}+R^{2}+2rR\cos\varphi\right)^{-\alpha/2}}\right)rdr\right),
\end{eqnarray}
where (a) follows from the probability generating functional of the
PPP.
\else
\begin{align}
L_{I_{2}}\left(s\right) & =  \mathbb{E}\left(\prod_{x\in\Phi_{\left[2\right]}}e^{-s\left(h_{x}l(x)+h_{m(x)}l(m(x))\right)}\right)\label{eq:I2-1}\\
 & =  \mathbb{E}\left(\prod_{x\in\Phi_{\left[2\right]}}\frac{1}{1+sl(x)}\frac{1}{1+sl(m(x))}\right)\label{eq:I2}\\
 & \overset{\left(a\right)}{=}  \exp\left(-\lambda p_{2}\int_{\mathbb{R}^{2}}\left(1-v(x)\right)\right)\nonumber\\
 & =  \exp\left(-\lambda p_{2}\int_{0}^{\infty}\left(1-\frac{K(s,r,R,\alpha)}{1+sr^{-\alpha}}\right)rdr\right)\nonumber
\end{align}
where (a) follows from the probability generating functional of the
PPP with $v(x)=\frac{1}{1+sl(x)}\frac{1}{1+sl(m(x))}.$
\fi
As a result, the success probability is 
\begin{align*}
p_{s} & =  L_{I_{1}}\left(\theta R^{\alpha}\right)L_{I_{2}}\left(\theta R^{\alpha}\right)\\
 & =  \exp\left(-\lambda p_{1}G(\theta R^{\alpha},\alpha)\right)\exp\left(-\lambda p_{2}F(\theta R^{\alpha},\alpha,R)\right),
\end{align*}
which completes the proof.
\end{IEEEproof}
The fact that the success probability (and the Laplace transform of the interference)
are a product of two terms follows from the
independence of the point processes $\Phi_{\left[i\right]}$.
The success probability is not in closed-form due to the integral form of $F(\theta R^{\alpha},\alpha,R)$. However,
tight bounds can be obtained.
\begin{thm}
The success probability is lower and upper bounded by
\[
\underline{p}_{s}=\exp\left(-\lambda\left(p_{1}+2p_{2}\right)G(\theta R^{\alpha},\alpha)\right)
\]
and
\[
\overline{p}_{s}=\exp\left(-\lambda\left(p_{1}+p_{2}\left(1+\delta\right)\right)G(\theta R^{\alpha},\alpha)\right).
\]
\end{thm}
\begin{IEEEproof}
Bounds only need to be established for the second term of the product in the success
probability. 

Lower Bound: From (\ref{eq:I2-1}),
\ifCLASSOPTIONonecolumn
\begin{eqnarray}
L_{I_{2}}\left(s\right) & = & \mathbb{E}\left(\prod_{x\in\Phi_{\left[2\right]}}e^{-s\left(h_{x}l(x)+h_{m(x)}l(m(x))\right)}\right)\nonumber \\
 & \overset{\left(a\right)}{\geq} & \mathbb{E}\left(\prod_{x\in\Phi_{\left[2\right]}}e^{-sh_{x}l(x)}\right)\mathbb{E}\left(\prod_{x\in\Phi_{\left[2\right]}}e^{-sh_{m(x)}l(m(x))}\right)\label{eq:lb2}\\
 & \overset{\left(b\right)}{=} & \exp\left(-2\lambda p_{2}G(s,\alpha)\right),\nonumber 
\end{eqnarray}
\else
\begin{align}
L_{I_{2}}\left(s\right) & =  \mathbb{E}\left(\prod_{x\in\Phi_{\left[2\right]}}e^{-s\left(h_{x}l(x)+h_{m(x)}l(m(x))\right)}\right)\nonumber \\
 & \overset{\left(a\right)}{\geq}  \mathbb{E}\left(\prod_{x\in\Phi_{\left[2\right]}}e^{-sh_{x}l(x)}\right)\mathbb{E}\left(\prod_{x\in\Phi_{\left[2\right]}}e^{-sh_{m(x)}l(m(x))}\right)\label{eq:lb2}\\
 & \overset{\left(b\right)}{=}  \exp\left(-\lambda p_{2}G(s,\alpha)\right)\exp\left(-\lambda p_{2}G(s,\alpha)\right)\nonumber \\
 & =  \exp\left(-2\lambda p_{2}G(s,\alpha)\right),\nonumber 
\end{align}
\fi 
where (a) follows from the FKG inequality \cite[Lemma 1]{Vaze11} since
both $\prod_{x\in\Phi}e^{-sh_{x}l(x)}$ and $\prod_{x\in\Phi}e^{-sh_{m(x)}l(m(x))}$
are decreasing random variables. In (\ref{eq:lb2}), the first term
is similar to the calculation of $L_{I_{1}}\left(s\right)$ with $\Phi_{\left[1\right]}$
replaced by $\Phi_{\left[2\right]}$ while in the second term, $m(\Phi_{\left[2\right]}) $ is a PPP with the same density of $\Phi_{\left[2\right]}$
due to the displacement theorem \cite[page 35]{Haenggi12book}. As a result, the two factors in (\ref{eq:lb2}) are equal, and
\begin{align*}
p_{s} 
 & \geq  L_{I_{1}}\left(\theta R^{\alpha}\right)\exp\left(-2\lambda p_{2}G(\theta R^{\alpha},\alpha)\right)
 = \underline{p}_{s}.
\end{align*}

Upper Bound: The upper bound can be obtained from the Cauchy-Schwarz inequality.
From (\ref{eq:I2}),
\begin{align*}
L_{I_{2}}\left(s\right) & =  \mathbb{E}\left(\prod_{x\in\Phi_{\left[2\right]}}\frac{1}{1+sl(x)}\frac{1}{1+sl(m(x))}\right)\\
 & \leq  \left\{ K_{1}(s,\alpha)K_{2}(s,\alpha)\right\} ^{\frac{1}{2}}
\end{align*}
which follows from the Cauchy-Schwarz inequality with $K_{1}(s,\alpha)=\mathbb{E}\left(\prod_{x\in\Phi_{\left[2\right]}}\frac{1}{\left(1+sl(x)\right)^{2}}\right)$
and $K_{2}(s,\alpha)=\mathbb{E}\left(\prod_{x\in\Phi_{\left[2\right]}}\frac{1}{\left(1+sl(m(x))\right)^{2}}\right)$. We have
\begin{align*}
K_{1}(s,\alpha)
 & =  \exp\left(-2\pi\lambda p_{2}\int_{0}^{\infty}\left(1-\frac{1}{\left(1+sr^{-\alpha}\right)^{2}}\right)rdr\right)\\
 & =  \exp\left(-\pi\lambda p_{2}\left(1+\delta\right)\Gamma(1+\delta)\Gamma(1-\delta)s^{\delta}\right)\\
 & =  \exp\left(-\lambda p_{2}\left(1+\delta\right)G(s,\alpha)\right).
\end{align*}
$K_{2}(s,\alpha)=K_{1}(s,\alpha)$ because $m(\Phi_{\left[2\right]})$
is a PPP with the same density as $\Phi_{\left[2\right]}$. As
a result, 
\begin{align*}
L_{I_{2}}\left(s\right) & \leq  \left\{ K_{1}(s,\alpha)K_{2}(s,\alpha)\right\} ^{\frac{1}{2}}\\
 & =  \exp\left(-\lambda p_{2}\left(1+\delta\right)G(s,\alpha)\right).
\end{align*}
Therefore, 
\begin{align*}
p_{s} 
 & \leq  e^{-\lambda p_{1}G\left(\theta R^{\alpha},\alpha\right)}e^{-\lambda p_{2}\left(1+\delta\right)G\left(\theta R^{\alpha},\alpha\right)}
 = \overline{p}_{s}.
\end{align*}
\end{IEEEproof}
The lower bound can be intuitively
understood as lower bounding the interference of the FD nodes (which
are formed by two dependent PPPs) by that of two independent PPPs with the same
density.

The upper bound turns out to be the same as the result
by assuming $l(x)=l(m(x))$ i.e., the distances between the receiver
at the origin and the interfering pair from the FD links are the same. 
Indeed, assuming $l(x)=l(m(x))$, we have\begin{align*}
\tilde{L}_{I_{2}}\left(s\right) & =  \mathbb{E}\left(\prod_{x\in\Phi_{\left[2\right]}}e^{-s\left(h_{x}+h_{m(x)}\right)l(x)}\right)\\
 & =  \exp\left(-\pi\lambda p_{2}\mathbb{E}\left[\left(h_{x}+h_{m(x)}\right)^{\delta}\right]\Gamma\left(1-\delta\right)s^{\delta}\right)\\
 & =  \exp\left(-\pi\lambda p_{2}\Gamma(2+\delta)\Gamma(1-\delta)s^{\delta}\right)\\
 & =  \exp\left(-\pi\lambda p_{2}\left(1+\delta\right)\Gamma(1+\delta)\Gamma(1-\delta)s^{\delta}\right)\\
 & =  \exp\left(-\lambda p_{2}\left(1+\delta\right)G(s,\alpha)\right)
\end{align*}
where $\mathbb{E}\left[\left(h_{x}+h_{y}\right)^{\delta}\right]=\Gamma(2+\delta)$
comes from the fact that $h_{x}+h_{y}$ has an Erlang distribution
and $\Gamma(2+\delta)=\left(1+\delta\right)\Gamma(1+\delta)$
is due to the property of the gamma function. Hence, the approximated
success probability assuming $l(x)=l(m(x))$ is $\tilde{p}_{s} = L_{I_{1}}\left(\theta R^{\alpha}\right)\tilde{L}_{I_{2}}\left(\theta R^{\alpha}\right)= \overline{p}_{s}$.

This result is not surprising. The equality holds for the Cauchy-Schwarz
inequality if $\prod_{x\in\Phi_{\left[2\right]}}\frac{1}{\left(1+sl(x)\right)^{2}}$
and $\prod_{x\in\Phi_{\left[2\right]}}\frac{1}{\left(1+sl(m(x))\right)^{2}}$
are linearly dependent. Obviously, $l(x)=l(m(x))$ satisfies this condition. 
Therefore, we have $\tilde{p}_{s}=\overline{p}_{s}$
as expected. 
Also, ${\overline{p}_{s}}/\underline{p}_{s}=e^{\lambda p_{2}\left(1-\delta\right)G\left(\theta R^{2},\alpha\right)}\rightarrow1$ as $\lambda\rightarrow0$ or $R\rightarrow0$.
Hence, the bounds are very tight in the asymptotic case. 
\vspace{0mm}
\begin{figure}[h]
\begin{centering}
\includegraphics[width=\figwidth]{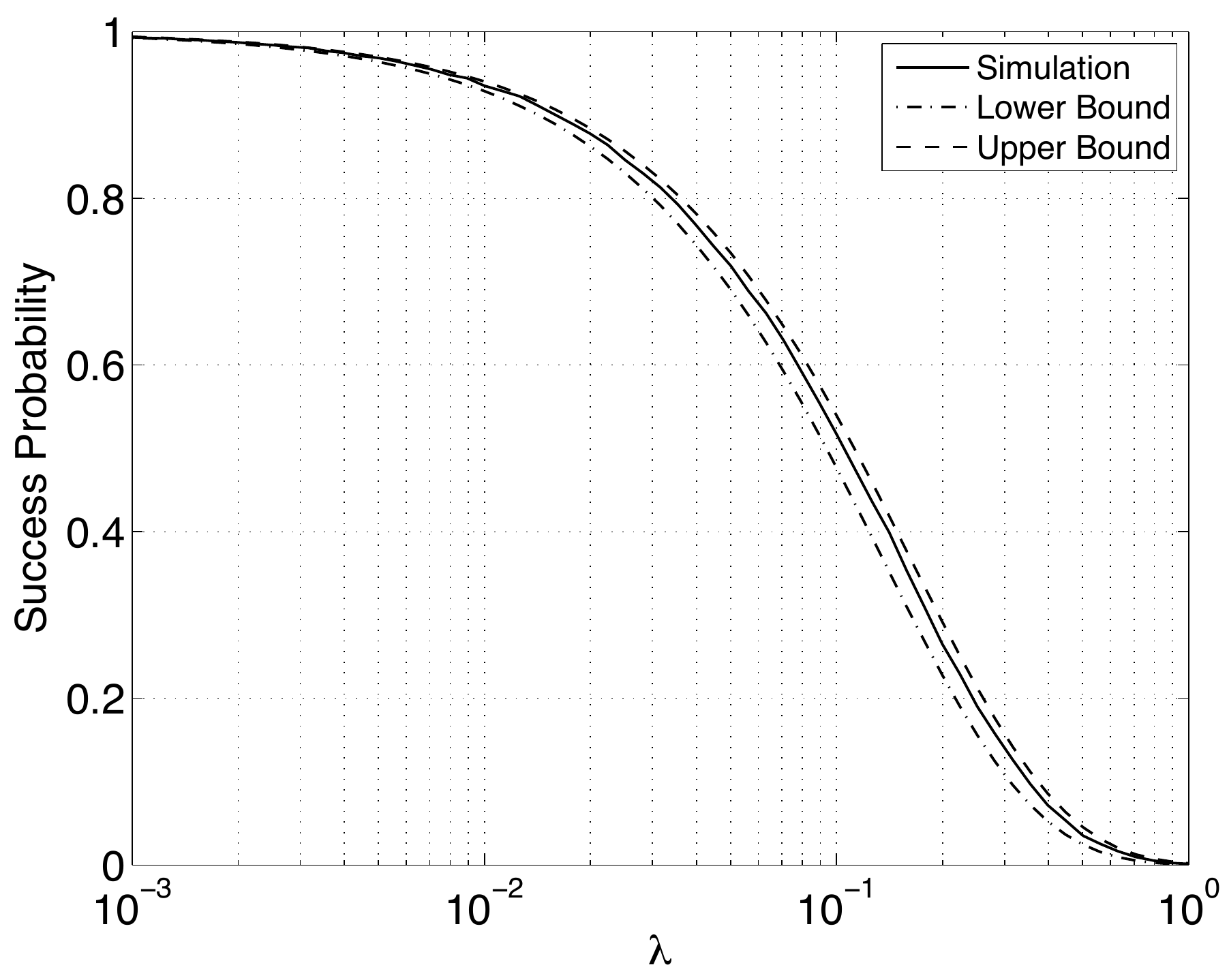}
\par\end{centering}

\caption{Comparison of success probability between simulation and its bounds
as a function of the node density $\lambda$: $\alpha=4$, $\theta=1$, $R=1$, 
$p_{0}=0$, $p_{1}=p_{2}=0.5$. }
\label{Fig:1}
\vspace{-3mm}
\end{figure}
\begin{sloppypar}
Figure \ref{Fig:1} plots the success probability from simulation
and its closed-form upper and lower bounds as a function of the node density.
As seen, both bounds are tight.
\begin{cor}
\label{cor:F}The function $F(s,\alpha,R)$ can be bounded as follows:
\[
\left(1+\delta\right)G(s,\alpha)\leq F(s,\alpha,R)\leq2G(s,\alpha).
\]
 \end{cor}
\begin{IEEEproof}
From the proof of the upper and lower bounds of the success probability,
we can easily derive the above inequalities.
\end{IEEEproof}
This corollary is useful in calculating the maximal throughput and
its bounds in the following section.

\section{Throughput Performance Analysis\label{sec:Throughput-Performance-Analysis}}
\subsection{Problem Statement}
The goal of FD transmission in a network is to increase
the network throughput. While FD increases the link throughput, it also causes additional interference to the other links. Given a network
that consists of nodes of FD and HD capabilities,
how should a node choose between FD and HD transmissions as the network configuration varies? Should
the node always transmit in a FD mode, or should it always work in
HD mode? Or should it sometimes work in FD mode while sometimes in
HD mode? It is unlikely to conclude to make the nodes work in HD mode
all the time. Otherwise, there is no need for FD. However, it is interesting to figure out which is better
between a FD-only network or a FD/HD-mixed network. In our model, the question
is equivalent to given a network that consists of nodes with both HD and FD capabilities,
how can we choose $p_{1}$ and $p_{2}$ to optimize the throughput
in the network? To see that, we first define the throughput. In a
random wireless network described by $\hat{\Phi}$, we can consider
the typical link, consisting of a node $x_{0}$ and its mark $m(x_{0})$.
The \textit{typical link} has probability $p_{1}$ to be in HD mode and
$p_{2}$ to be in FD mode. Therefore, its throughput can be defined as follows:
\vspace{0mm}
\begin{defn}
For a wireless network described by $\hat{\Phi}$, the throughput
of the typical link is defined as
\begin{equation}
T=\left(p_{1}+2p_{2}\right)p_{s}\label{eq:T}
\end{equation}

\end{defn}
Inserting $p_{s}$ from (\ref{eq:ps-2}) into (\ref{eq:T}), we have
\ifCLASSOPTIONonecolumn
\begin{equation}
T\left(p_{1},p_{2}\right)=\left(p_{1}+2p_{2}\right)\exp\left(-\lambda p_{1}G\right)\exp\left(-\lambda p_{2}F\right).\label{eq:T1}
\end{equation}
\else
\begin{equation}
T\left(p_{1},p_{2}\right)=\left(p_{1}+2p_{2}\right)e^{-\left(\lambda p_{1}G+\lambda p_{2}F\right)}.\label{eq:T1}
\end{equation}
\fi 

From now on, we will
use $G$ to denote $G(\theta R^{\alpha},\alpha)$ and $F$
to denote $F(\theta R^{\alpha},\alpha,R)$ for simplicity.
Given the definition of throughput, there are two extreme cases that
are particularly interesting: one is the case where all concurrently
transmitting nodes work in HD mode, i.e., $p_{2}=0,$ and the other
is where all concurrently transmitting nodes work in FD mode, i.e.,
$p_{1}=0$. Their throughputs are given as:
$
T^{\mbox{\scriptsize{HD}}}(p_{1})=p_{1}\exp\left(-\lambda p_{1}G\right)
$
and
%
$
T^{\mbox{\scriptsize{FD}}}(p_{2})=2p_{2}\exp\left(-\lambda p_{2}F\right).
$
\vspace{0mm}
\subsection{Throughput Optimization}

It is interesting to find the relationship between the maximal values
of $T^{\mbox{\scriptsize{HD}}}$, $T^{\mbox{\scriptsize{FD}}}$and $T$, denoted as $T_{\max}^{\mbox{\scriptsize{HD}}}$,
$T_{\max}^{\mbox{\scriptsize{FD}}}$ and $T_{\max}$. In other words, we would
like to see how to choose $p_{1}$ and $p_{2}$ such that the maximal
throughput of the network is achieved. 
First of all, $T_{\max}^{\mbox{\scriptsize{HD}}}$ and $T_{\max}^{\mbox{\scriptsize{FD}}}$
can be easily obtained by the following lemma.
\begin{lem}
\label{lem:6}For a wireless network of HD-only network, described
by $\hat{\Phi}$ with $p_{2}=0$, $T_{\max}^{\mbox{\scriptsize{HD}}}$ is given
by 
\begin{equation}
T_{\max}^{\mbox{\scriptsize{HD}}}=\begin{cases}
T^{\mbox{\scriptsize{HD}}}(\frac{1}{\lambda G})=\frac{1}{\lambda G}e^{-1} & \mbox{if }\lambda G\geq1\\
T^{\mbox{\scriptsize{HD}}}(1)=e^{-\lambda G} & \mbox{if }\lambda G<1
\end{cases},\label{eq:hdmax}
\end{equation}
with optimal MAP 
\vspace{0mm}
\begin{equation}
p_{1}^{{\mbox{\scriptsize{opt}}}}=\min\left(\frac{1}{\lambda G},1\right).\label{eq:p1opt}
\end{equation}
 Similarly, for a wireless network of FD-only network, described by
$\hat{\Phi}$ with $p_{1}=0$, $T_{\max}^{\mbox{\scriptsize{FD}}}$ is given by
\begin{equation}
T_{\max}^{\mbox{\scriptsize{FD}}}=\begin{cases}
T^{\mbox{\scriptsize{FD}}}(\frac{1}{\lambda F})=\frac{2}{\lambda F}e^{-1} & \mbox{if }\lambda F\geq1\\
T^{\mbox{\scriptsize{FD}}}(1)=2e^{-\lambda F} & \mbox{if }\lambda F<1
\end{cases},\label{eq:fdmax}
\end{equation}
with optimal MAP 
\vspace{-4mm}
\begin{equation}
p_{2}^{{\mbox{\scriptsize{opt}}}}=\min\left(\frac{1}{\lambda F},1\right).\label{eq:p2opt}
\end{equation}
\end{lem}
\begin{IEEEproof}
The proof is straightforward by taking the derivatives of $T^{\mbox{\scriptsize{HD}}}$
and $T^{\mbox{\scriptsize{FD}}}$ with respect to $p_{1}$ and $p_{2}$.
\end{IEEEproof}
In the following theorem, we show that $T_{\max}$ is achieved by
setting all concurrently transmitting nodes to be in FD mode and that $T_{\max} = T_{\max}^{\mbox{\scriptsize{FD}}}$. 
\begin{thm}
For a wireless network described by $\hat{\Phi}$, the maximal throughput
is given by 
\begin{equation}
T_{\max}=T_{\max}^{\mbox{\scriptsize{FD}}},\label{eq:equ}
\end{equation}
with the optimal MAP $\left(p_{1},p_{2}\right)=\left(0,p_{2}^{\mbox{\scriptsize{opt}}}\right)=\left(0,\min\left(\frac{1}{\lambda F},1\right)\right)$.
Also, (\ref{eq:equ}) holds regardless of the network configuration
$\left(\lambda,\theta,R,\alpha\right)$.\end{thm}
\begin{IEEEproof}
Taking derivative of $T$ w.r.t. $p_{1}$ and $p_{2}$ leads to 
\vspace{-3mm}
\begin{equation}
\frac{\partial T}{\partial p_{1}}=\exp\left(-\lambda\left(p_{1}G+p_{2}F\right)\right)\left[1-\lambda G\left(2p_{2}+p_{1}\right)\right],\label{eq:p1}
\end{equation}
\vspace{-3mm}
\begin{equation}
\frac{\partial T}{\partial p_{2}}=\exp\left(-\lambda\left(p_{1}G+p_{2}F\right)\right)\left[2-\lambda F\left(2p_{2}+p_{1}\right)\right].\label{eq:p2}
\end{equation}
Note that $2p_{2}+p_{1}\in\left[0,2\right]$ and $\lambda F<2\lambda G$
from Corollary \ref{cor:F}.
\leavevmode
\begin{enumerate} \item $\lambda F<2\lambda G<1$: $2\lambda G<1$ leads to $\frac{\partial T}{\partial p_{1}}>0$. Therefore, $T$ is an increasing function in $p_{1}$. $\lambda F<1$ implies $\frac{\partial T}{\partial p_{2}}>0$. $T$ is also an increasing function in $p_{2}$. As a result, $T_{\max}=\max\left\{ T^{\mbox{\scriptsize{HD}}}(1),T^{\mbox{\scriptsize{FD}}}(1)\right\} $ . Since  \begin{eqnarray*} \frac{T^{\mbox{\scriptsize{FD}}}(1)}{T^{\mbox{\scriptsize{HD}}}(1)} = 2e^{\lambda\left(G-F\right)} \overset{\left(a\right)}{\geq} 2e^{-\lambda F/2} \overset{\left(b\right)}{\geq} 2e^{-\frac{1}{2}} > 1, \end{eqnarray*} where (a) follows from Corollary  \ref{cor:F} and (b) from $\lambda F<1$, $T_{\max}=T^{\mbox{\scriptsize{FD}}}(1)=T^{\mbox{\scriptsize{FD}}}_{\max}$ in this case.  \item $\lambda F<1<2\lambda G$: Under this condition, let $\frac{\partial T}{\partial p_{1}}=0$ and we have $2p_{2}+p_{1}=1/{\lambda G}.$ Also, $\frac{\partial T}{\partial p_{2}}>0$ still holds. Therefore, the maximal $T$ is achieved at $\left(p_{1},p_{2}\right)=\left(0,\frac{1}{2\lambda G}\right)$ from $2p_{2}+p_{1}=\frac{1}{\lambda G}$ and $\frac{\partial T}{\partial p_{2}}>0$. Note that $T(0,\frac{1}{2\lambda G})=T^{\mbox{\scriptsize{FD}}}(\frac{1}{2\lambda G})<T^{\mbox{\scriptsize{FD}}}(1)$. Hence, $T_{\max}=T^{\mbox{\scriptsize{FD}}}(1)=T^{\mbox{\scriptsize{FD}}}_{\max}$. \item $1<\lambda F<2\lambda G$: Let $\frac{\partial T}{\partial p_{1}}=0$ and $\frac{\partial T}{\partial p_{2}}=0$ and we have  \begin{equation} 2p_{2}+p_{1}=1/{\lambda G}\label{eq:dp1} \end{equation}  and  \begin{equation} 2p_{2}+p_{1}=2/{\lambda F}.\label{eq:dp2} \end{equation} From (\ref{eq:dp1}), $T$ is maximized at $p_{1}=\frac{1}{\lambda G}-2p_{2}$, which leads to  \begin{eqnarray*} T = \frac{1}{\lambda G}e^{\lambda p_{2}\left(2G-F\right)-1}   \leq  T(0,\frac{1}{2\lambda G}) =  \frac{1}{\lambda G}e^{-\frac{F}{2G}}\end{eqnarray*} From (\ref{eq:dp2}), $T$ is maximized at $p_{2}=\frac{1}{\lambda F}-\frac{p_{1}}{2}$, which leads to \begin{eqnarray*} T  =  \frac{2}{\lambda F}e^{\lambda p_{1}\left(\frac{F}{2}-G\right)-1}   \leq  T(0,\frac{1}{\lambda F})=  \frac{2}{\lambda F}e^{-1} \end{eqnarray*} Therefore, $T_{\max}=\max\left\{ T(0,\frac{1}{2\lambda G}),T(0,\frac{1}{\lambda F})\right\} $. Since  \begin{eqnarray*} \frac{T(0,\frac{1}{2\lambda G})}{T(0,\frac{1}{\lambda F})}  = \frac{F}{2G}e^{1-\frac{F}{G}} \overset{\left(a\right)}{\leq} e^{1-\frac{F}{G}}\overset{\left(b\right)}{\leq}  e^{-\delta} < 1, \end{eqnarray*} where (a) follows from $F<2G$ and (b) from $F>\left(1+\delta\right)G$. Hence, $T_{\max}=T^{\mbox{\scriptsize{FD}}}(\frac{1}{\lambda F})=T^{\mbox{\scriptsize{FD}}}_{\max}$.\end{enumerate}\end{IEEEproof}
To summarize, 
\[
T_{\max}=T_{\max}^{\mbox{\scriptsize{FD}}}
\]for all the cases, which means $T_{\max}$ is always achieved by
setting all transmitting nodes to work in FD mode instead
of in HD mode, i.e. $p_{1}=0$, despite the interference
caused by the FD nodes. This conclusion is not affected by the network
configuration $(\lambda, \theta, R, \alpha)$. The corresponding optimal
MAPs is $\left(p_{1}^{\mbox{}},p_{2}^{\mbox{}}\right)=\left(0,\min\left(\frac{1}{\lambda F},1\right)\right)$,
which is equivalent to setting $p_{2}$ to be the optimal MAP of FD-only
network given in (\ref{eq:p2opt}) with $p_{1}=0$.
%
\end{sloppypar}

\subsection{Comparison of FD with HD}
\begin{figure}[h]
\begin{centering}
\includegraphics[width=\figwidth]{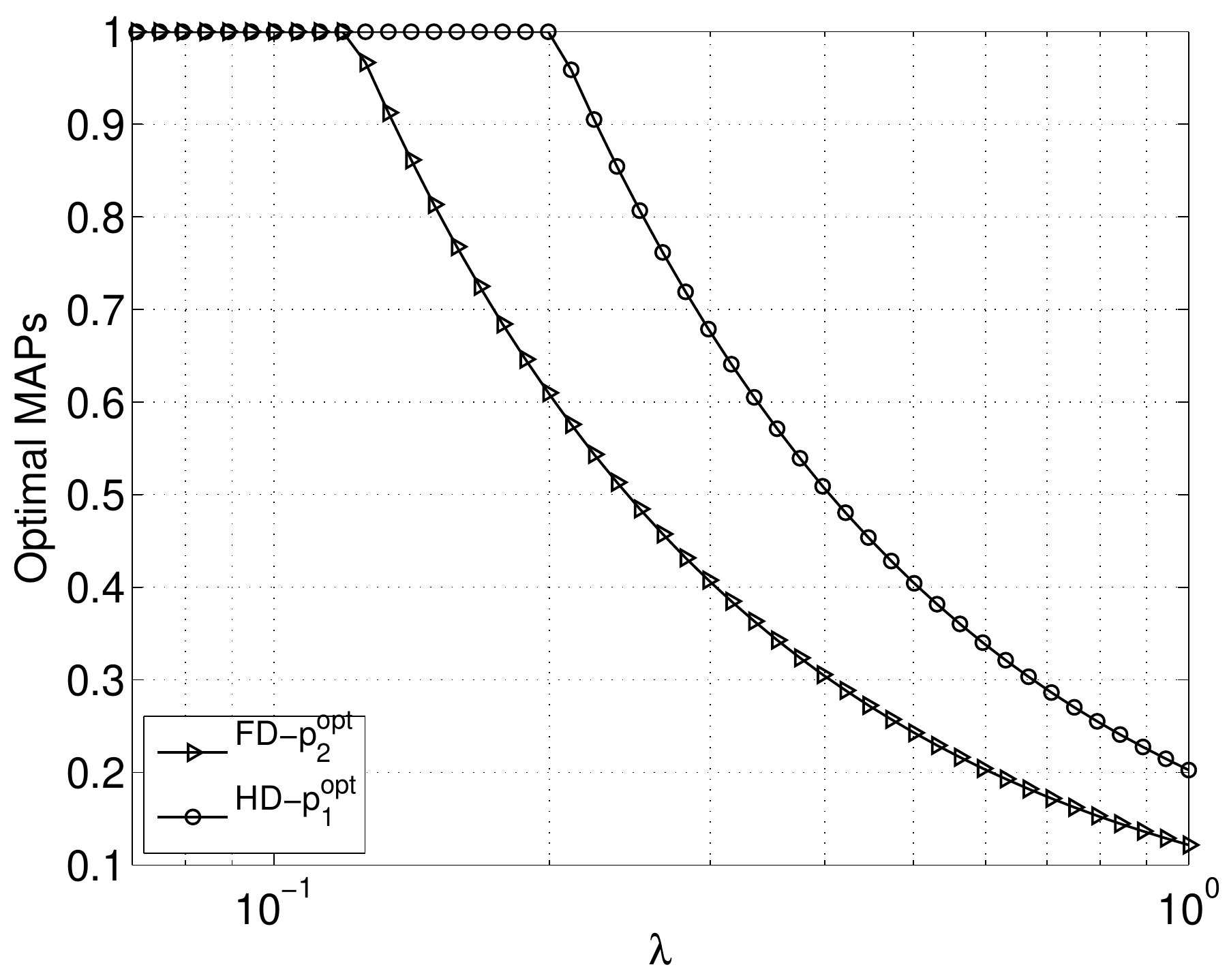}
\par\end{centering}
\caption{Optimal MAPs that achieve the maximal throughput for FD and HD networks
as a function of the node density $\lambda$: $\alpha=4$, $R=1$, $\theta=1$.}
\label{Fig:3}
\vspace{-3mm}
\end{figure}

Since the mixed FD/HD network will achieve maximal throughput at
the extreme case of an FD-only network, we can simply focus on the FD-only
and HD-only networks and compare their optimal MAPs and maximal throughputs
from the results in Lemma \ref{lem:6}. Given a fixed set of system
parameters $\left(\theta,R,\alpha\right)$, the optimal MAPs to achieve
the maximal throughput for FD and HD networks as a function of node
density $\lambda$ are illustrated in Figure \ref{Fig:3}. The figure
is plotted according to (\ref{eq:p1opt}) and (\ref{eq:p2opt}) and
it shows that both FD and HD network will make all nodes transmit when
the node density is lower than $\frac{1}{F}$ and
$\frac{1}{G}$, respectively, and after that the MAPs will be inversely
proportional to the node density $\lambda$. 
The throughput gain of FD network with respect to HD network is of
great interest. In the following, the throughput gain is defined:
\begin{defn}
The throughput gain ($\rm{TG}$) is defined as the ratio between the maximal
throughput of FD network and HD network given the same network configuration
$\left(\lambda,\theta,R,\alpha\right)$:
\[
\rm{TG}=\frac{T_{\max}^{\mbox{\scriptsize{FD}}}}{T_{\max}^{\mbox{\scriptsize{HD}}}}.
\]
\end{defn}
The following corollary gives the theoretical expression of $\rm{TG}$ in
terms of $\lambda$, $F$ and $G$ together with its lower and upper
bounds in closed-form. Note that $F$ and G are constant given fixed
$\left(\theta,R,\alpha\right)$.
\begin{cor}
The throughput gain is given by
\vspace{-3mm}
\begin{equation}
\rm{TG}=\begin{cases}
2e^{\lambda\left(G-F\right)} & \mbox{if }\lambda F<1,\lambda G<1\\
\frac{2}{\lambda F}e^{\left(\lambda G-1\right)} & \mbox{if }\lambda F\geq1,\lambda G<1\\
\frac{2G}{F} & \mbox{if }\lambda F\geq1,\lambda G\geq1
\end{cases}\label{eq:TG}
\end{equation} and bounded as
\begin{equation}
\begin{cases}
2e^{-\lambda G}  < \rm{TG} < 2e^{-\delta\lambda G}  & \mbox{if }\lambda F<1,\lambda G<1\\
\frac{e^{\left(\lambda G-1\right)}}{\lambda G} <  \rm{TG} < \frac{2e^{\left(\lambda G-1\right)}}{\left(1+\delta\right)\lambda G} & \mbox{if }\lambda F\geq1,\lambda G<1\\
1 < \rm{TG} < \frac{2}{1+\delta} & \mbox{if }\lambda F\geq1,\lambda G\geq1
\end{cases}\label{eq:TG_ub}
\end{equation}
\end{cor}
\begin{IEEEproof}
From (\ref{eq:hdmax}) and (\ref{eq:fdmax}), when $\lambda F<1$
(it implies $\lambda G<1$ from Corollary \ref{cor:F}), $\rm{TG}={T^{\mbox{\scriptsize{FD}}}(1)}/{T^{\mbox{\scriptsize{HD}}}(1)}=2e^{\lambda\left(G-F\right)}$;
when $\lambda F>1$ and $\lambda G<1$, $\rm{TG}={T^{\mbox{\scriptsize{FD}}}({1}/{\lambda F})}/{T^{\mbox{\scriptsize{HD}}}(1)}={2}e^{\left(\lambda G-1\right)}/{\lambda F}$;
and when $\lambda F>1$ and $\lambda G>1$, $\rm{TG}={T^{\mbox{\scriptsize{FD}}}({1}/{\lambda F})}/{T^{\mbox{\scriptsize{HD}}}({1}/{\lambda G})}={2G}/{F}$.
Therefore, we have (\ref{eq:TG}). The upper and lower bounds can
be easily proven by using Corollary \ref{cor:F}. Note that $G=G(\theta R^{\alpha},\alpha)$
and hence both bounds are in closed form.
\end{IEEEproof}
\vspace{-3mm}
\begin{figure}[h]
\begin{centering}
\includegraphics[width=\figwidth]{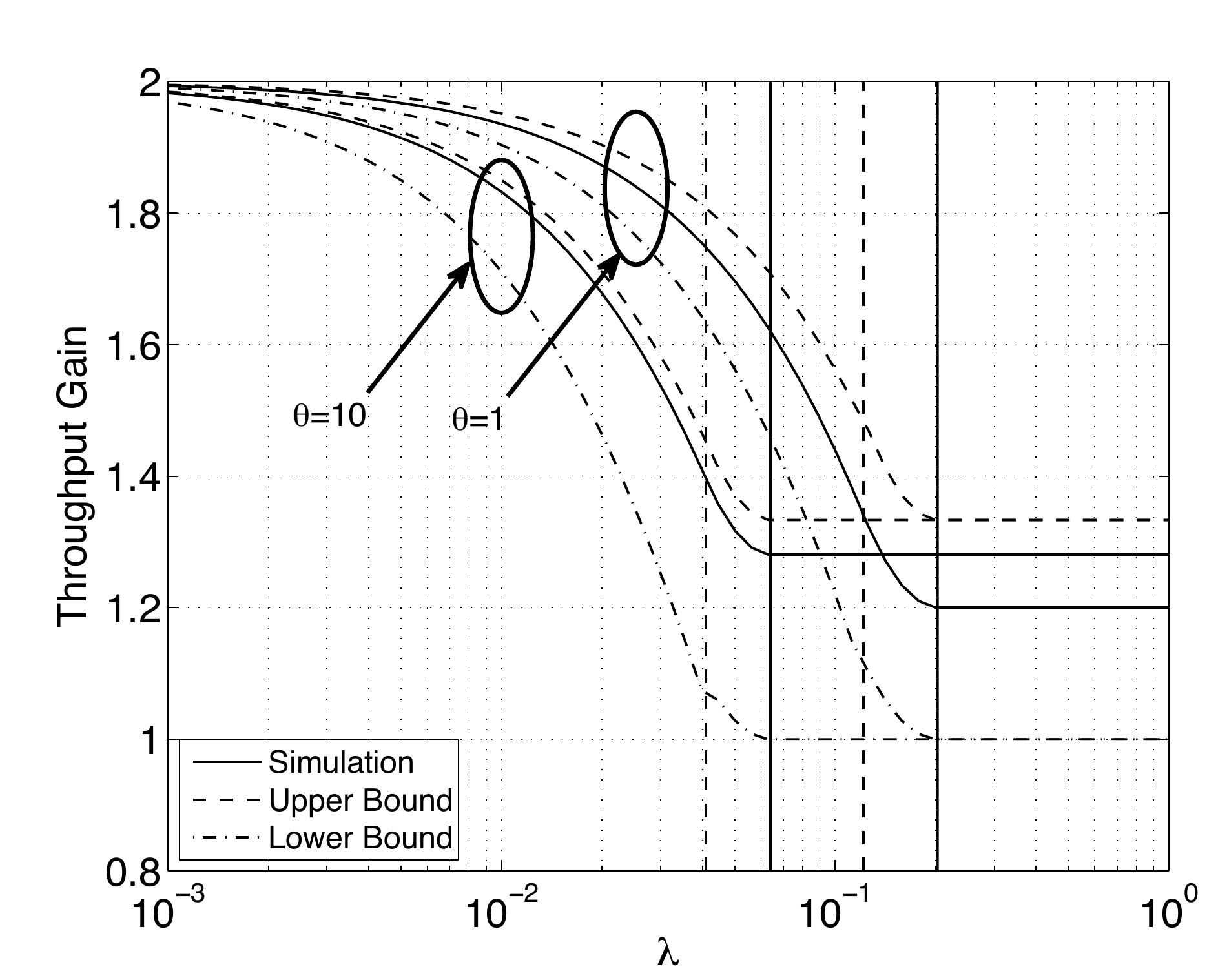}
\par\end{centering}
\caption{Throughput gain as a function of the node density $\lambda$ and its bounds:
$\alpha=4$, $R=1$. The two dashed vertical lines indicate the point where $\lambda= 1/F$ for different $\theta$ while the two solid vertical lines indicate $\lambda = 1/G$. Given $\theta$, $R$, and $\alpha$, the estimate of $F(\theta R^{\alpha},\alpha,R)$ can be obtained
numerically. }
\label{Fig:4}
\vspace{-3mm}
\end{figure}
 Figure \ref{Fig:4} illustrates the throughput gain as a function
of the node density together with its upper and lower bounds given in (\ref{eq:TG_ub}). As seen, the throughput gain
is approaching $2$ asymptotically as the node density $\lambda\rightarrow0$.
As the node density increases, the throughput gain decreases. The throughput gain will become constant
after the node density is greater than the threshold $1/G$, which is illustrated by two solid vertical lines for different $\theta$. The reason is that the density of concurrent transmitting nodes will get saturated under ALOHA protocol for both FD and HD networks if $\lambda$ is great than $1/G$. This constant
throughput gain is upper bounded by $\frac{2}{1+\delta}$ and lower
bounded by $1$. The effect of the SINR threshold $\theta$ is that a larger
SINR threshold gets the network saturated at lower node density since a larger SINR threshold implies the system can tolerate less interference. As a result, the constant throughput gain will be higher for larger $\theta$. 
In general, an ALOHA protocol will guarantee
the throughput gain of the FD transmission to be greater than $1$
for all network configurations, which means that FD transmission
always outperforms HD transmission.

\section{Conclusion\label{sec:Conclusion}}

In this paper, we analyzed the throughput of wireless networks with
FD radios using mathematical tools from stochastic geometry. Given a wireless network of radios
with both FD and HD capabilities, we showed that FD transmission is always preferable
compared to HD transmission in terms of throughput. Although the throughput of FD transmission can not be doubled,
the gain is considerable in an ALOHA protocol. In general, FD can be a
very powerful technique that can be adapted for the next-generation
wireless networks if the limitations from real world can be tackled, i.e., the imperfect self-interference cancellation.
Moreover, the throughput gain is expected to be larger if more advanced
MAC protocols other than ALOHA are used or the interference management
can be used for the pairwise interferers in the FD links.
%
\bibliographystyle{IEEEtran}
\bibliography{reference}

\end{document}